\documentclass[sigconf]{acmart}
\usepackage{rotating, multirow, multicol, csquotes}
\usepackage{arydshln, booktabs, threeparttable}
\AtBeginDocument{%
  }

\setcopyright{acmlicensed}
\copyrightyear{2025}
\acmYear{2025}
\acmConference[ICAIL '25]{Twentieth International Conference on Artificial Intelligence and Law}{June 16--20,
  2025}{Chicago, IL}

\usepackage{xspace}
\PassOptionsToPackage{frozencache,cachedir=_minted-main}{minted} 
\usepackage{minted}
\usepackage{tabularx}
\usepackage{bm}
\usepackage{hyperref}
\usepackage{subcaption}
\usepackage{afterpage}
\usepackage{tikz}
\usepackage{pgfplots}
\pgfplotsset{compat=1.17} 

\newcommand{\benchmark}{\texttt{LaborBench}\xspace}
\newcommand{\corpus}{\texttt{StateCodes}\xspace}

\definecolor{systemcolor}{RGB}{44, 130, 201}
\definecolor{usercolor}{RGB}{40, 167, 69}
\definecolor{assistantcolor}{RGB}{255, 87, 51}
\usepackage{placeins}
\usepackage{titlesec}
\titleformat{\subsection}[runin]
  {\normalfont\normalsize\bfseries}
  {\thesubsection}{1em}{}
  [.]

\begin{document}

\title{AI for Statutory Simplification: A Comprehensive State Legal Corpus and Labor Benchmark}

\author{Emaan Hariri}
\email{ehariri@stanford.edu}
\affiliation{%
  \institution{Stanford University}
  \city{Stanford}
  \state{CA}
  \country{USA}
}

\author{Daniel E. Ho}
\email{dho@law.stanford.edu}
\affiliation{%
  \institution{Stanford University}
  \city{Stanford}
  \state{CA}
  \country{USA}
}

\renewcommand{\shortauthors}{Hariri and Ho}

\begin{abstract}
One of the emerging use cases of AI in law is for code simplification: streamlining, distilling, and simplifying complex statutory or regulatory language. One U.S.\ state has claimed to eliminate one third of its state code using AI. Yet we lack systematic evaluations of the accuracy, reliability, and risks of such approaches. We introduce \benchmark, a question-and-answer benchmark dataset designed to evaluate AI capabilities in this domain. We leverage a unique data source to create \benchmark: a dataset updated annually by teams of lawyers at the U.S.\ Department of Labor, who compile differences in unemployment insurance laws across 50 states for over 101 dimensions in a six-month process, culminating in a 200-page publication of tables. Inspired by our collaboration with one U.S. state to explore using large language models (LLMs) to simplify codes in this domain, where complexity is particularly acute, we transform the DOL publication into \benchmark. This provides a unique benchmark for AI capacity to conduct, distill, and extract realistic statutory and regulatory information. To assess the performance of retrieval augmented generation (RAG) approaches, we also compile \corpus, a novel and comprehensive state statute and regulatory corpus of 8.7 GB, enabling much more systematic research into state codes. We then benchmark the performance of information retrieval and state-of-the-art large LLMs on this data and show that while these models are helpful as preliminary research for code simplification, the overall accuracy is far below the touted promises for LLMs as end-to-end pipelines for regulatory simplification.\footnote{\corpus and \benchmark are available \url{https://huggingface.co/collections/reglab/a-reasoning-focused-legal-retrieval-benchmark-67a00c363f7e0d14619e95c5}.}
\end{abstract}





\maketitle

\section{Introduction} \label{sec:intro}

While AI is increasingly integrated into legal practice---using AI to research positive law ---one of the emerging, ambitious, and potentially most consequential use cases lies in legal reform itself---using AI to normatively change the law. An area of increasing political focus has been statutory and regulatory reform. Given the vast terrain of statutory and regulatory provisions, can advances in AI and large language models (LLMs), help with code and regulatory simplification, revision, and reform?  Many politicians, AI vendors, and researchers have espoused this use case. After cutting some 22k royal decrees, Italy's Minister for Institutional Reform touted ``AI for regulatory simplification'' to identify duplicative, conflicting, or outdated requirements that can trap citizens and ``clog up the courts,'' \cite{redazione_di_rainews_casellati_2023, aptusai_too_2023}. The U.S. Department of Health and Human Services used AI in its ``regulatory cleanup initiative'' to  identify outdated or erroneous provisions \cite{HHS}. Relying on an AI tool marketed by Deloitte, the state of Ohio has purported to eliminate 5M words (or one third) of its state code, with Ohio Lieutenant Governor Husted noting ``an unprecedented opportunity to use AI tools to eliminate regulations, to make them more easily understandable, and thus make them easier to comply with,'' \cite{lips_ohio_2024}. The benefits for governments, citizens, and civil servants could be immense. Former Deputy Chief Technology Officer Jen Pahlka, who helped found the US Digital Service, said, ``[i]f you use AI for that regulatory simplification \dots you’ll be burdening your folks less, so that they can do more of the right things that you care about on behalf of the people in your state,'' \cite{jenkins_lawmakers_2024}. Most ambitiously, the Trump Administration's regulatory reform agenda has unveiled the Elon Musk-led ``Department of Government Efficiency'' (DOGE) as a reorganization of the US Digital Service to streamline regulatory requirements using technology across all regulatory agencies (with reported utilization of AI to recommend revisions to regulations  \cite{gilbert_doge_2025}). Within the private sector, vendors have touted tools for regulatory simplification (e.g., Deloitte's RegExplorer, Aptus.AI) and a wide range of ``RegTech'' firms have integrated AI to automatically scan for diverse and changing regulatory requirements across jurisdictions and for simplifying compliance initiatives (see, e.g., \cite{karanam_radically_2020}).

What is sorely lacking, however, are systematic evaluations of these capacities. To date, legal benchmarks have focused on discrete questions of rule retrieval, application, and reasoning. The ambition of AI for regulatory reform, however, requires systematic benchmarking of \emph{legal code interpretation} of statutes and regulations far beyond existing benchmarks. 

We hence introduce \benchmark, a new benchmark dataset to assess the capacity of state-of-the-art models for code understanding, and \corpus, a comprehensive dataset of nearly all underlying state statutory codes and regulations, totaling 5.1 and 3.6 GB, respectively.  Our contributions are sixfold. 

First, we leverage a unique source of data. Every few years, a team of a half dozen lawyers at the U.S.\ Department of Labor spends roughly six months combing through complex state statutes and regulations---described as ``byzantine'' \cite{harris_byzantine_2021} and requiring ``a master’s degree in confusion''  --to formally characterize differences in state unemployment insurance (UI) schemes.  Their efforts result in a 200-page compilation, entitled ``Comparison of State Unemployment Insurance Laws,'' of tabular comparisons of key code differences along hundreds of dimensions across the states. This benchmark is a unique and highly realistic one---requiring years of effort by teams of lawyers and subject matter experts that continue annually as provisions are revised.  We turn this into a high-fidelity dataset  of 3,700 questions and answers in \benchmark. 

Second, this setting is a highly realistic and consequential one for code simplification. It was inspired by a direct collaboration with a state to explore the use of LLMs to simplify unemployment insurance (UI). UI is one of the largest benefits programs to provide cash payments to workers who lose their job through no fault of their own. During the pandemic, roughly 46M Americans relied on UI, but the complexity of benefits determinations has led to strong calls for simplifying the UI thicket. In the book ``Recoding America,'' Pahlka devotes an entire chapter to the ``new guy,'' who despite 17 years of experience was still learning the ropes of benefits determinations \cite{Pahlka2023-li}. The annual DOL compilation forms an important component of federal oversight into a federal-state cooperative scheme. More generally, many legal AI providers have marketed their AI capacity to conduct 50-state surveys to capture differences in laws across jurisdictions.\footnote{The surveys can be either \textit{cross-sectional}, focusing one point in time, or \textit{longitudinal}, looking at multiple points in time \cite{burris_technical_2014}.}  Thomson Reuters, for instance, markets Westlaw with the ability to ``creat[e] comprehensive and up-to-date surveys of the law using generative AI.''\cite{thomson_reuters_generative_2024} \citet{burris_technical_2014} estimates that a thoroughly researched manual survey might take 3 months, with more complex areas of law requiring even more time and substantial teams. 

Third, one of the emerging techniques in legal AI has been the use of retrieval augmented generation (RAG). Unfortunately, comprehensive, structured sources of all U.S. state statutes and regulations are not easily available to researchers, impeding systematic research in legal AI. We hence introduce \corpus, which includes all statutory code from 50 U.S. states\footnote{``State'' in this paper will be used to refer to state-level jurisdictions, including the territories and the federal district (D.C.).}, containing 488M words across 1.8M sections, and state regulations from 45 U.S. states, containing 268M words across 2.2M sections. We use this dataset to test RAG type approaches to solving the \benchmark tasks, but this dataset provides an ideal setting to develop further 50-state RAG benchmarks in other areas of law.

Fourth, we benchmark five state-of-the-art models (GPT-4o mini, Gemini 1.5 Flash, Claude 3.5 Haiku, Llama 3.1 70B Instruct, and Deepseek V3) with retrieval augmentation using a range of dense and sparse retrieval methods (E5, Gemini, Okapi BM25, OpenAI) on \benchmark. We find that while LLMs may help in preliminary research for code simplification, the overall accuracy is far below the touted promises for LLMs as end-to-end pipelines for regulatory simplification (F1 of 0.67). To provide one simple example, UI laws define employers as persons who employed individuals on 20 days across different weeks in a calendar year. But complexities in statutory language led Gemini identify the threshold for Colorado's statute as \textit{20 days}, while erroneously identifying the threshold of the federal model statute as \textit{20 weeks}\footnote{When we exam the statutes, it is apparent they are similarly worded. \textit{Compare} \textsc{Colo. Rev. Stat.} § 8-70-113(1)(a)(II)(B) (``Employed at least one individual in employment for some portion of the day on each of twenty days during the calendar year \dots each day being in a different calendar week.'') \textit{with} 26 U.S.C. § 3306(a)(1)(B) (``[O]n each of some 20 days during the calendar year \dots each day being in a different calendar week, employed at least one individual in employment for some portion of the day.'')\label{fn:colo_futa_example}}. These mistakes are not isolated. Overall accuracy of the best performing model (Gemini) is 0.68, demonstrating that there are significant advances required for applications in code simplification. 

Fifth, we show, however, that AI systems may nonetheless aid in the human process. We design a set of experiments to test for the impact of information retrieval (IR) in a RAG pipeline. This allows us to separately investigate (a) the performance on the IR component (citation verification), and (b) the performance of models without an IR component, which replicate the substantive setting of a search without and with knowing the relevant code provisions, respectively.  We find that RAG approaches boost performance substantially, increasing F1 by 0.18, demonstrating the need for the \corpus dataset and improved IR methods.  Moreover, we show that citation elicitation via chain-of-thought prompting, even if imperfect (top-5 recall of 0.93), can aid in legal research process for code simplification. 

Last, one of the unique aspects of \benchmark is that state statutes and regulations follow from a cooperative federal-state scheme, with a model federal statute. Longstanding questions exist around state legislative and regulatory capacity \cite{lindsey2021state}. We illustrate how we can examine how AI assistance interacts with such capacity differences with the Correlates of State Policy dataset \cite{grossmann2021correlates}. 

This paper proceeds as follows. Section~\ref{sec:related} discusses related works that motivate the design of \benchmark and \corpus. Section~\ref{sec:datasets} articulates the construction of our new benchmark. Section~\ref{sec:experiment} discusses our evaluation approach and experimental setup. Section~\ref{sec:results} presents results and Section~\ref{sec:conclusion} concludes with brief implications. 

\section{Related Works}\label{sec:related}

Efforts at improving legal retrieval and reasoning have spanned decades \cite{locke_case_2022, meldman_preliminary_1975}. We explore recent work focuses on applying newer retrieval methods and domain-specific legal QA. 

\subsection{Statute Retrieval} Our work contributes to important work on statutory retrieval.  
The Competition on Legal Information Extraction/Entailment (COLIEE) \cite{goebel_overview_2024} has been important for focusing community efforts on legal AI and contains an information retrieval task (Task 3) and an entailment/QA task (Task 4). These tasks use the Official English Translation of the Japanese Civil Code, with 768 Articles, as their corpus. Task 3 involves retrieving the appropriate subset of articles for a query $Q$--- appropriate meaning those articles that are dispositive of $Q$---and contains 996 test and 101 test queries. Task 4 involves the same queries datasets, but evaluates only the entailment task, i.e., ``Yes/No'' question answering, rather than retrieval itself. While insights from COLIEE have advanced the understanding of retrieval in legal reasoning, it remains a relatively small corpus sourced from one jurisdiction.  \benchmark builds on these efforts by providing a benchmark for large-scale multi-jurisdictional retrieval and analysis.

\citet{su_stard_2024} developed STAtute Retrieval Dataset (STARD), a Chinese-language dataset consisting of 1,543 queries  lacking precise legal terminology made by \textit{non-professionals}. The corpus used is comprised of 55,348 statutory articles selected from the statutory and regulatory law of China. The stated objective is to retrieve the smallest set of relevant statutes necessary to answer the query, where relevancy was determined by law student annotators. The best method tested by \citet{su_stard_2024} achieves a recall@100 of 0.907. STARD provides valuable insights on non-professional queries in one jurisdiction; our work expands on this by including technical queries that cover 50 U.S. states on the same legal topic. 

Other datasets provide large sets of queries but have relatively small corpora and limited target statutes. The Chinese AI Law (CAIL) dataset created by \citet{xiao_cail2018_2018} focuses on legal judgment prediction from 2.6 million criminal cases, filtered to include only those arising from 183 articles of the Chinese Criminal Code. \citet{paul_legal_2024} introduced the Indian Legal Statute Identification (ILSI) and the European Court of Human Rights (ECtHR-B) datasets, the latter building on \citet{chalkidis_neural_2019}'s ECHR dataset. ILSI contains 65,950 queries derived from Indian Supreme Court cases, with 100 target statutes from the Indian Penal Code, while ECtHR-B has 10,825 queries, with just 10 target statutes from the European Convention on Human Rights, all in English. While ECtHR-B focuses on distinct statutes, ILSI highlights challenges in retrieving and reasoning over similarly worded laws\footnote{\citet{paul_legal_2024}  compare Indian Penal Code (India Act XLV), 1860, §§ 323, 325, 326 with Convention for the Protection of Human Rights and Fundamental Freedoms arts. 2–14, Nov. 4, 1950, 213 U.N.T.S. 222 as an example.}. Our work extends this focus by providing an order of magnitude more target statutes and addressing similar legal provisions across diverse jurisdictions\footnote{For instance, \benchmark includes QAs about UI definitions, which are often similar. \textit{Compare, e.g.}, \textsc{Mass. Gen. Laws} ch. 151A, § 1(j) \textit{with} \textsc{Hawaii Rev. Stat.} § 381-1\label{fn:ui_defs}{c}.}, enabling richer evaluation of LLMs' handling of semantic overlap and nuanced statutory language.

\subsection{State Law Corpora} One of the major impediments to research on U.S. state jurisdictions has been the lack of a structured corpus on state statutes and regulations.  \citet{henderson_pile_2022}, for instance, include U.S. state codes in the Pile of Law, but codes are concatenated and then chunked into 150 sections evenly and Pile of Law does not contain state regulations, which often play an outsized role in legal implementation. \corpus is the first dataset we are aware that offers U.S. state codes in a structured manner and contains the vast majority of U.S. state regulations.

\subsection{Legal RAG Benchmarks}

Existing benchmarks address a wide array of legal reasoning tasks, including legal judgment prediction, topic classification, interpretation,  knowledge memorization, and issue spotting
\cite{chalkidis_lexglue_2022, guha_legalbench_2023, fei_lawbench_2023}. These benchmarks, however, place less emphasis on the retrieval and processing of very long texts, which is increasingly important given the rise of legal RAG systems \cite{magesh_hallucination-free_2024}. \citet{guha_legalbench_2023} introduced LegalBench, which encompasses 162 tasks comprehensively covering six types of legal reasoning but excludes tasks that focus on processing long documents. \citet{fei_lawbench_2023} explicitly exclude from their benchmark, LawBench, any tasks like legal case retrieval that involve reasoning over long documents. LegalBench contains only 9 sections of text (sourced from the Internal Revenue Code) exceeding 5K tokens. \corpus provides a highly realistic benchmark, with extensive context length, with, for instance, over 100 sections exceeding 5K tokens. 

\citet{pipitone_legalbench-rag_2024} developed LegalBench-RAG to focus specifically on the evaluation of RAG systems in the legal domain. LegalBench-RAG contains 6,858 queries which are divided so as to relate to 4 different corpora---NDA related documents, private contracts, public M\&A documents, and privacy policies---containing 79M characters. This benchmark covers exclusively private law matters, not including case law, statutory, or regulatory retrieval tasks. With 4.8B characters, \corpus expressly focuses on the important setting of statutory and regulatory law. 

\section{Datasets} \label{sec:datasets}

We now describe the creation of \corpus---state statutes and regulations across all domains---and \benchmark, which enables assessment of legal RAG in the domain of unemployment insurance (UI). Focusing on UI has several benefits. First, UI is notoriously complex \cite{Pahlka2023-li}, providing an ideal testing ground. Second, all state laws nonetheless must address the same minimum set of issues, due to the cooperative federal scheme. Many provisions are modeled on the Federal Unemployment Tax Act (FUTA) (as illustrated in Footnotes~\ref{fn:colo_futa_example} and ~\ref{fn:ui_defs}). This structure enables the tabular comparisons compiled by DOL that we transform into \benchmark. Third, UI resembles many other areas of law where states rely heavily on model statutes to promote legal consistency across jurisdictions, while retaining flexibility to tailor provisions.\footnote{For instance, the Uniform Commercial Code is at least partially adopted in all 50 states and D.C.\textsc{U.C.C.} (\textsc{Am. Law Inst. \& Unif. Law Comm'm}); Within estate law, the Uniform Probate Code see wide usage. \textsc{Unif. Prob. Code}, 8 pt. 1 U.L.A. 1 (2013).} The value of \corpus beyond testing just \benchmark is clear, as it enables researchers to systematically study state statutory and regulatory law across all domains.

\begin{figure*}[!t]
    \centering
    \includegraphics[width=\linewidth]{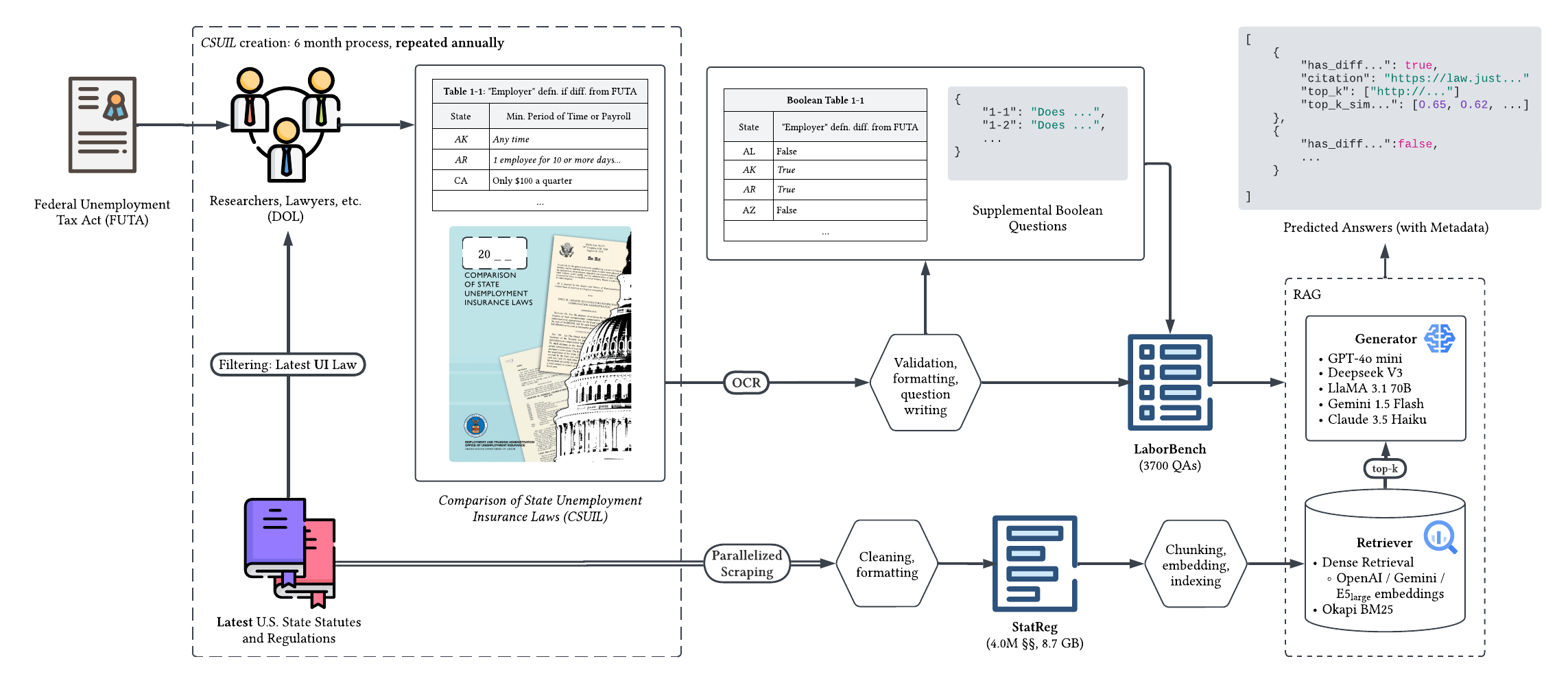}
    \caption{Summary of compilation and benchmarking processes of \benchmark and \corpus.} \label{fig:splash}
\end{figure*}

For a simplified illustration of the steps involved in creating and benchmarking \benchmark and \corpus, see Figure \ref{fig:splash}.

\subsection{\benchmark}

\subsubsection{Source Data.}

The data underlying \benchmark was sourced from the 2023 edition of the \textit{Comparison of State Unemployment Insurance Laws} (\textit{CSUIL}), an annually published report which describes itself as providing ``state-by-state information on workers covered, benefit eligibility, methods of financing, and other areas of interest in the Unemployment Insurance'' \cite{schuettinger_comparison_2023}. 

As noted, many legal experts spend almost six months updating \textit{CSUIL} annually. In developing state surveys, lawyers define legal questions; identify divergence from federal legislation; and update table accuracy based on recent legislation, regulation, and case law.  

The report is the compilation of 86 extensive state comparison tables spread across 8 different chapters, each representing a subfield of UI. Each state survey is a topic-focused table where the rows represent states and the columns each represent a subtopic related to the topic. The majority of the tables have 53 rows, representing all 50 states, the District of Columbia, and the territories of Puerto Rico and the U.S. Virgin Islands (although some tables have fewer rows, with one having only 2 rows). The average table has justf over 2 columns. Each column will have entries of a common data type, which may be either string, integer, or boolean;  many string columns in the dataset may be cast into categorical columns (e.g., some tables in \textit{CSUIL} chapter 5 have columns akin to \texttt{[`L', `R', `L', `I', `I',...]}). 

Some table entries contain footnotes. When storing these entries in the dataset, the footnote is removed and the footnote text is saved under a separate `metadata' key for that entry. Additionally, there is a master table of abbreviations covering the table.


\subsubsection{Question/Answer (QA) Construction.}\label{sec:qa_construct} We employ the tables to construct a QA dataset with (1) each table's respective context, usually the text immediately preceding each table, and (2) an auxiliary set of questions derived from the headers of the original table. 

For example, \textit{CSUIL} Table 3-16, entitled ``Withholding State and Local Income Tax at Claimant Option'' and having the column  ``State Taxes,'' yields the following question: 
\begin{displayquote}
    Given the description above, can state income taxes be withheld from unemployment insurance benefits at the claimant's option in \{jurisdiction\}?
\end{displayquote}
For each row of the table, we replace the placeholder text `jurisdiction' with the full jurisdiction (typically state) name.

Each entry of \benchmark is composed of: (1) an index (integer), (2) the column number (integer), (3) the column name (string written in snake case (e.g. `has\_different\_employer\_definition')), (4) the column data type (string, one of `bool', `str', or `int'), (5) the jurisdiction (string), (6) the abbreviation for that jurisdiction (string), (7) question context (string), (8) a question derived from the column header and jurisdiction (string), and (9) the answer to that question (type corresponding to the column data type).

\subsection{Boolean Supplement to \benchmark}

We also construct a boolean supplement (with binary answers). We start with the initial set of boolean questions from \benchmark and add 1,272 questions, derived from the state survey tables. For each of 20 tables, we typically formulate a question that asks whether the state has express statutory language on the subject. For simplicity, we use one-hot encodings for categorical answers. This query can then be formulated to \textit{all} jurisdictions. For example, \textit{CSUIL} Table~1-1 (in Table~\ref{fig:table_nonbool_example}(a)) lists provisions defining ``employer'' in states where the definition deviates from the FUTA. From this, we generate Table~\ref{fig:table_nonbool_example}(b), a boolean table indicating for all states whether their definitions deviate from FUTA. 


\begin{table}[htb]
    \caption{Converting (a) \textit{CSUIL} Table 1-1, Definition of Employer (If Different from FUTA 20 Weeks/\$1,500 Rule), to (b) a Boolean Table for \benchmark} \label{fig:table_nonbool_example}
    \centering
    \begin{subtable}[t]{0.45\linewidth}
        \centering
        \caption{}
        \begin{tabular}{c|l}
            \toprule
            State & Min. Time/Payroll$^*$ \\ 
            \midrule
            AK & Any time \\
            AR & 1 employee \dots \\
            CA & Over \$100 \dots \\
            \vdots & \vdots \\
            \bottomrule
        \end{tabular}
        \label{tab:nonbool_a}
    \end{subtable}
    \hfill 
    \begin{subtable}[t]{0.45\linewidth}
        \centering
        \caption{}
        \begin{tabular}{c|c}
            \toprule
            State & FUTA Diff.$^\dagger$ \\
            \midrule
            AL &  \\
            AK & X \\
            AR & X \\
            AZ &  \\
            \vdots & \vdots \\
            \bottomrule
        \end{tabular}
        \label{tab:nonbool_b}
    \end{subtable}
    \vspace{0.5em}
    \par\raggedright \small 
    $^*$ Minimum Period of Time or Payroll \\
    $^\dagger$ Defines ``Employer'' Differently From FUTA
\end{table}

\subsection{\corpus}

In addition to the QAs, we provide 50 U.S. state statutory codes and 45 U.S. state regulatory codes (or administrative codes) in a structured corpus. This corpus dates to 2023, corresponding to \benchmark's source data (\textit{CSUIL} 2023). Sections of statute and regulation exhibit great variation in length: statutory code sections have average of 265 words ($\textrm{SD}=817$), while regulatory code sections have an average of 122 words ($\textrm{SD}=780$). The most verbose statutory and regulatory sections have 332K and 609K words, respectively. Different state statutory codes vary in size from having 4,700 code sections (Illinois) to nearly 160K sections (California).

We source our state statutes and regulations from Justia, a legal information provider.\footnote{To handle the structure and sheer size of the state laws, we write a script which collects them in parallel. This script can be used to collect the laws for any given year. This script can be found here: \url{https://github.com/reglab/lawscraper}.} The U.S. regulations on Justia are themselves sourced from Fastcase \cite{justia_us_2024}.\footnote{We note that such governmental materials, including state statutes and regulations, are not subject to copyright protection. \textit{See generally} Georgia~v. Public.Resource.Org Inc., Inc., 590 U.S. 255 (2020)); Wheaton~v. Peters, 33 U.S. 591 (1834).}  While state codes and regulations can be obtained digitally from each state, there is little standardization in the way the states offer their codes digitally  \cite{california_legislature_california_2025, texas_legislature_texas_2024, washington_state_legislature_revised_2024}). Even within the same state, codes and regulations may not be standardized \cite{california_legislature_california_2025, california_office_of_administrative_law_california_2025}.  

While much legal AI has focused on case law research, state statutes and regulations are of particular interest, as they determine how the largest government programs are administered, exhibit distinct language, and have extensive scope and complexity. 
Table~\ref{fig:statute_sizes} provides a sense of the size of \corpus relative to existing corpora and benchmarks. 

\begin{table}[h]
    \caption{Relative Size of Statutory Corpora Used for IR or LLM Benchmarking} \label{fig:statute_sizes}
    \begin{tabular}{c|cc}
        \toprule
         Legal Corpus&Characters&Sections\\
         \midrule
         COLIEE (Task 3) \cite{goebel_overview_2024}&$\sim$315K&768 \\
         CAIL \cite{xiao_cail2018_2018}&113K& 183\\ 
         STARD \cite{su_stard_2024}&6.6M& 55K \\
         ILSI \cite{paul_legal_2024}&$\sim$40K&100 \\ 
         U.S.\ Code&49M&54K+\\
         \corpus (Statutes) &3.0B&1.8M\\
         \corpus (Regs.) &1.8B&2.2 M\\
         \bottomrule
    \end{tabular}
\end{table}

\section{Experimental Setup} \label{sec:experiment}

We focus our experiments on the boolean subset of \benchmark, using a range of generative and retrieval models with a filtered subset of \corpus of UI laws.\footnote{We determine which sections of each jurisdiction's statutory and regulatory codes correspond to UI laws. Hand labeling is required because states publish their UI laws under different names. \textit{See, e.g.}, \textsc{Kan.\ Stat.\ Ann.} §§ 44-701 to 44-777  (``Employment Security Act''); \textsc{Fla.\ Stat.} XXXI.443 (``Reemployment Assistance''). This is done using a regular expression filter over the citation URLs.} This allows us to retain their complexity while making our analysis more tractable, but researchers can of course make the task more difficult by omitting this filter. (As we show below, this task proves sufficiently difficult for the range of systems we consider.) Table~\ref{ref:sample_run} presents an example of prompts and system interaction. 

\FloatBarrier
\begin{table*}[htb]
    \footnotesize
    \caption{Annotated Sample Evaluation of \benchmark using LLM}\label{ref:sample_run}
    \centering
    \begin{tabular}{@{}p{0.075\textwidth} p{0.7\textwidth} p{0.225\textwidth}@{}}
    \toprule
    \textbf{Role} & \textbf{Content} & \textbf{Step} \\
    \midrule
    \textcolor{systemcolor}{\textbf{System}} &
    \raggedright
    Given the context, generate a response to the user query:\newline
    \begin{tabular}{@{}p{0.125\linewidth}p{0.875\linewidth}@{}}
    \textit{Document 1:} & 2023 Alaska Statutes › \dots › Art. 9. General Provisions. › Sec. 23.20.520. Definitions. \\
    \textit{Cite URL 1:} & \texttt{https://.../alaska/title-23/chapter-20/article-9/section-23-20-520/} \\
    \textit{Context 1:} & In this chapter, unless the context otherwise requires: (1) ``American vessel'' means \dots \newline (2) ``average annual wage'' means \dots (10) ``employer'' means \dots 
    \end{tabular}\newline
    \dots \newline
    Explain your reasoning, and think step by step \dots &
    \textbf{0.}  Context
    is retrieved from \corpus using $Q$. \\
    \midrule
    \textcolor{usercolor}{\textbf{User}} &
    \raggedright $\bm{Q}$: ``Does Alaska define employer differently than FUTA?'' &
     \textbf{1.} Query $Q\in\benchmark$ using above system prompt. \\
    \textcolor{assistantcolor}{\textbf{Assistant}} &
    \raggedright
    Yes, the state of Alaska defines ``employer'' differently from the FUTA (Federal Unemployment Tax Act) standards, particularly regarding the 20 weeks/\$1,500 rule. Under the Alaska Employment \dots &
     \\
    \textcolor{usercolor}{\textbf{User}} &
    \raggedright
    Please respond using the following JSON schema: 
    \begin{minted}{javascript}
{"properties": {
    "has_different_employer_definition": { "type": "boolean", ..., }
    "citation_url": ...
  }, ...}
    \end{minted} 
    &
    \textbf{2.} Request the answer conform to the JSON Schema derived from $Q$. \\
    \textcolor{assistantcolor}{\textbf{Assistant}} & \vspace{-1em}
\begin{minted}[escapeinside=||]{javascript}
{"has_different_employer_definition": |\colorbox{green}{true}|,
"citation_url": ".../alaska/title-23/chapter-20/article-9/section-23-20-520/"}
    \end{minted} 
    \vspace{-4em}
    &
    (\textbf{3.} Result may be validated using JSON Schema.) \\
    \bottomrule
    \end{tabular}
\end{table*}

\subsection{RAG}

Legal RAG has gained in popularity, given its effectiveness on knowledge-intensive NLP tasks like open-domain QA \cite{lewis_retrieval-augmented_2021} and ability to reduce hallucinations \cite{shuster_retrieval_2021}. \benchmark and its boolean supplement provide an ideal setting for testing RAG using \corpus. We test a number of retrieval methods, including dense passage retrieval (DPR) \cite{lewis_retrieval-augmented_2021, karpukhin_dense_2020} and sparse retrieval, against a baseline. 

We test DPR using three popular embedding models, two proprietary one open source. The proprietary models we use are OpenAI's `text-embedding-3-small' \cite{openai_new_2024} and Google's `text-embedding-004' \cite{lee_gecko_2024}. We also use the E5\textsubscript{large} model (`e5\_large\_v2') from the open source family of E5 models\cite{wang_text_2024}, itself derived from BERT-based MiniLM\cite{wang_minilm_2020}. For sparse retrieval, we use the efficient and popular Okapi BM25 retrieval model \cite{robertson1995okapi, Manning2008-dw, jones2000probabilistic, robertsonProbabilisticRelevanceFramework2009}.  

\subsection{Ingestion}
To ingest \corpus, we implement a hybrid of semantic and fixed-size chunking strategies, since \corpus (both its statutes and regulations) are organized into individual sections. Here we choose to ingest both corpora, but we may choose to ingest just one (or neither, in the case of our baseline) of them. In particular, chunking is done only \textit{within} each statute section, i.e., a chunk will never span two sections. Within each section, fixed-size chunking is used with a chunk size of 1,000 tokens and an overlap of 200 tokens.\footnote{While chunking is not strictly necessary for Okapi BM25, long documents have been shown to degrade Okapi BM25's retrieval performance \cite{lv_when_2011}.} We use the GPT-4 tokenizer (tiktoken with encoding `cl100k\_base') at this stage. 

Top-$k$ retrieval for relevant chunks is done using Maximum Inner Product Search (MIPS) for DPR. Upon querying, MIPS is used to obtain the top-$k$ relevant \textit\textit{chunks}. For context selection, we then obtain the full \textit{section} (from \corpus) associated with each chunk to provide more context to the model during inference.\footnote{It is possible many of the top-$k$ chunks retrieved are from the same section. In this case, we drop the chunks associated with lower similarities (since the resulting set of retrieved \textit{sections} will be the same).} For the purposes of our experiments, we use $k=5$.

Our query is augmented in a straightforward manner.  For each inference model text (see below) the sections derived from the top-$5$ chunks are placed in the system prompt in this fashion:
{
\small
\begin{verbatim}
    Given the context, generate a response to the user query: 
    Document 1: {title[0]}
    Citation URL: {url[0]}
    Context: {content[0]}
    Document 2: {title[1]}
    ...
\end{verbatim}
}

Where  \texttt{title[i]} is the title of the \texttt{i}\textsuperscript{th} section of statute or regulation fetched (of the modified top-$k$, i.e., $0 \le i < k$), which contains the path,\footnote{For instance, a full title looks like ``2023 Code of Alabama › Title 1 - General Provisions. › Chapter 1 - Construction of Code and Statutes. › Section 1-1-1 - Meaning of Certain Words and Terms.''}, \texttt{url[i]} is the Justia URL linking to that statute or regulation, and \texttt{content[i]} is the text of that section. 

\subsection{Inference}\label{sec:infer}

For generation, we tested five popular LLMs: GPT-4o mini (gpt-4o-mini-2024-07-18), Gemini 1.5 Flash, Claude 3.5 Haiku (claude-3-5-haiku-20241022), Llama 3.1 70B Instruct, and the recently released Deepseek V3 \cite{deepseek-ai_deepseek-v3_2024}. Claude 3.5 Haiku and Llama 3.1 70B Instruct were only used with the OpenAI embedding model. To optimize the extraction of boolean responses when testing \benchmark's boolean supplement, we implemented a two-step inference strategy. 

First, the query from \benchmark was presented to the models, generally prompting them to provide their reasoning alongside the response. Optionally, we enhanced this step by explicitly requesting chain-of-thought (CoT) reasoning \cite{wei_chain--thought_2023, kojima_large_2023} and, where relevant, an explicit citation to support the response.

Second, providing all information from the previous step, we then prompt the model to produce a JSON fitting a schema. To dynamically create JSON Schemas for each query, we use a Pydantic model that contains at most fields: the first field has a name matching the column name for that query (e.g., `has\_different\_employer\_definition', see Section \ref{sec:qa_construct}) and a type matching the column data type (in the case of \benchmark's supplement, this will always be `bool'); the second and optional field is named `citation\_url' and is of type string. This allows us to obtain a final output that is boolean in the vast majority of cases.\footnote{Fewer than 20 out of over 20,000 results did not produce a boolean.}

\section{Results} \label{sec:results}

We now present results, which suggest  
that while LLMs may be useful aids, much effort is still needed to advance the performance of LLMs and retrieval models in this space.

\subsection{Retriever and Generator Model Performances}
The evaluation reveals notable differences in performance across various retriever and generator models, as highlighted in Table~\ref{tab:results_embed_model}. 

The Gemini retriever demonstrates strong performance across all metrics. It outperforms all other retrieval methods on all metrics except recall, see Table~\ref{tab:results_embed_model} (although the Gemini-GPT configuration has the highest recall among all RAG configurations). We observe that the BM25 generally lags behind the DPR methods in most metrics, albeit not substantially (and it even achieves higher precision than E5\textsubscript{large}). The BM25-Gemini configuration performs poorly across metrics.

Overall, the highest F1 revealed among all RAG configurations was 0.691, indicating that the problem remains challenging, forming a good benchmark task for measuring advances in this space. Even in the simplified boolean QA task with RAG, models misinterpret statutory and regulatory requirements over one third of the time.

\FloatBarrier
\begin{table*}[h]
    \caption{Comparison of Mean Performance Metrics Across Retriever and Generator Models ($\pm$ SE, calculated using a nonparametric bootstrap with $n=1000$)} \label{tab:results_embed_model}
    \centering
    \begin{tabular}{ll|cccc}
        \toprule
        \textbf{Retriever} & \textbf{Generator} & \textbf{Accuracy} & \textbf{Precision} & \textbf{Recall} & \textbf{F1} \\ 
        \midrule
        \multirow{4}{*}{Baseline ($\varnothing$)} & Deepseek & 0.534 ± 0.012 & 0.461 ± 0.014 & 0.673 ± 0.018 & 0.547 ± 0.013 \\ 
        ~ & Gemini & 0.564 ± 0.012 & 0.419 ± 0.037 & 0.114 ± 0.012 & 0.179 ± 0.018 \\ 
        ~ & GPT & 0.511 ± 0.012 & 0.450 ± 0.014 & 0.758 ± 0.014 & 0.564 ± 0.013 \\ 
        \cdashline{2-6}
        ~ & \textit{Average} & 0.536 ± 0.007&0.452 ± 0.010&0.515 ± 0.011&0.481 ± 0.009 \\
        \midrule
        \multirow{4}{*}{E5\textsubscript{large}} & Deepseek & 0.648 ± 0.012 & 0.555 ± 0.017 & 0.860 ± 0.014 & 0.675 ± 0.014 \\ 
        ~ & Gemini & 0.486 ± 0.041 & 0.426 ± 0.048 & 0.767 ± 0.055 & 0.548 ± 0.048 \\ 
        ~ & GPT & 0.625 ± 0.014 & 0.536 ± 0.017 & 0.866 ± 0.013 & 0.662 ± 0.014 \\ \cdashline{2-6} ~ & \textit{Average} & 0.628 ± 0.010&0.538 ± 0.012&0.858 ± 0.010&0.661 ± 0.010 \\ 
        \midrule
        \multirow{3}{*}{Gemini} & Deepseek & 0.674 ± 0.012 & 0.579 ± 0.017 & 0.857 ± 0.015 & \textbf{0.691 ± 0.014} \\ 
        ~ & Gemini & \textbf{0.716 ± 0.011} & \textbf{0.665 ± 0.021} & 0.670 ± 0.017 & 0.667 ± 0.015 \\ 
        ~ & GPT & 0.645 ± 0.014 & 0.550 ± 0.017 & \textbf{0.901 ± 0.013} & 0.683 ± 0.015 \\
        \cdashline{2-6}
        ~ & \textit{Average} & 0.679 ± 0.008&0.588 ± 0.011&0.809 ± 0.010&0.681 ± 0.009\\
        \midrule
        \multirow{4}{*}{Okapi BM25} & Deepseek & 0.635 ± 0.016 & 0.545 ± 0.019 & 0.847 ± 0.017 & 0.663 ± 0.017 \\ 
        ~ & Gemini & 0.705 ± 0.015 & 0.659 ± 0.023 & 0.632 ± 0.021 & 0.645 ± 0.019 \\ 
        ~ & GPT & 0.603 ± 0.015 & 0.519 ± 0.017 & 0.880 ± 0.015 & 0.653 ± 0.015 \\
        \cdashline{2-6}
        ~ & \textit{Average}&0.648 ± 0.008&0.561 ± 0.011&0.786 ± 0.010&0.654 ± 0.009 \\
        \midrule
        \multirow{5}{*}{OpenAI} & Claude & 0.692 ± 0.014 & 0.604 ± 0.019 & 0.798 ± 0.018 & 0.688 ± 0.016 \\ 
        ~ & Deepseek & 0.638 ± 0.013 & 0.547 ± 0.017 & 0.854 ± 0.013 & 0.667 ± 0.014 \\ 
        ~ & Gemini & 0.699 ± 0.014 & 0.654 ± 0.022 & 0.616 ± 0.021 & 0.635 ± 0.018 \\ 
        ~ & GPT & 0.650 ± 0.014 & 0.555 ± 0.018 & 0.889 ± 0.014 & 0.683 ± 0.015 \\ 
        ~ & Llama & 0.600 ± 0.014 & 0.517 ± 0.015 & 0.866 ± 0.015 & 0.647 ± 0.014 \\
        \cdashline{2-6}
        ~ & \textit{Average} & 0.656 ± 0.006&0.566 ± 0.008&0.805 ± 0.008&0.665 ± 0.007 \\
        \bottomrule
    \end{tabular}
\end{table*}

Table~\ref{fig:results_model} also provides a comparison of the average results by generator LLM (conditional on retrieval). 
Gemini exhibited relatively strong accuracy and recall results, but lower F1 scores. Claude and Llama had the highest and lowest F1 scores among generators, but were only tested with the OpenAI retriever; Deepseek had the highest F1 scores of those LLMs tested with all retrievers. Nevertheless, all F1 scores were within close range and leave substantial room for improvement.

\begin{table}[H]
    \caption{Comparison of Mean Performance Metrics Across Generator LLMs, Using Retrieval} \label{fig:results_model}
    \centering
    \begin{tabular}{l|cccc}
        \toprule
        \textbf{Generator} & \textbf{Accuracy} & \textbf{Precision} & \textbf{Recall} & \textbf{F1} \\ 
        \midrule
        Claude & 0.692 & 0.604 & 0.798 & \textbf{0.688} \\ 
        Deepseek & 0.649 & 0.556 & 0.855 & 0.674 \\ 
        GPT & 0.631 & 0.540 & \textbf{0.884} & 0.670 \\ 
        Gemini & \textbf{0.698} & \textbf{0.644} & 0.644 & 0.644 \\ 
        Llama & 0.600 & 0.517 & 0.866 & 0.647 \\ 
        \bottomrule
    \end{tabular}
\end{table}

\subsection{Measuring the Effects of RAG and CoT}\label{sec:rag_vs_baseline}

\subsubsection{Effects of RAG.}
Table~\ref{fig:rag_vs_no_rag} shows that, as expected \cite{lewis_retrieval-augmented_2021, karpukhin_dense_2020}, RAG provides a substantial performance increase across all relevant metrics. RAG is particularly effective at increasing recall, reducing the high number of false negatives with baseline LLMs. In Table \ref{tab:results_embed_model}, the baseline Gemini model appears to be an outlier, and RAG improves recall substantially. 

\begin{table}[h!]
    \caption{Performance Comparison of RAG and Baseline LLMs} \label{fig:rag_vs_no_rag}
    \begin{tabular}{l|cccc}
        \toprule
        \textbf{Method} & \textbf{Accuracy} & \textbf{Precision} & \textbf{Recall} & \textbf{F1} \\
        \midrule
        Baseline & 0.536  & 0.452 & 0.515 & 0.481 \\
        RAG & \textbf{0.656} & \textbf{0.567} & \textbf{0.806}  & \textbf{0.665} \\ 
        \bottomrule
    \end{tabular}
\end{table}

The asymmetric impact of retrieval on precision and recall suggests that while RAG helps models identify when statutory differences exist, it does not correspondingly improve their ability to either determine correctness or significantly reduce false positives. LLMs without retrieval struggle with false negatives, and RAG effectively and unsurprisingly addresses this issue by providing a needed knowledge source. However, the relatively small improvement in precision indicates that models remain limited in assessing the validity of retrieved statutes.

\subsubsection{Effects of Citation Elicitation Using Chain of Thought.} 

We use chain-of-thought (CoT) to invoke reasoning in the model \cite{lewis_retrieval-augmented_2021} while simultaneously producing a citation, selected from the retrieved (top-5) sources. Specifically, we apply CoT while requesting the generator explicitly provide a ``relevant'' citation for its answer (while allowing it to provide no answer if all citations appear irrelevant). Relative to RAG,  citation elicitation with CoT appears to have little effect on performance. While performance improves to a degree as shown in Table \ref{fig:cot_vs_no_cot}, most increases are minimal. 


\begin{table}[htb!]
    \caption{Performance Impacts of Citation Elicitation (CE) with Chain of Thought (CoT) over Baseline} \label{fig:cot_vs_no_cot}
    \begin{tabular}{l|cccc}
        \toprule
        \textbf{Method} & \textbf{Accuracy} & \textbf{Precision} & \textbf{Recall} & \textbf{F1} \\
        \midrule
        Baseline & 0.656  & 0.567 & \textbf{0.806}& 0.665 \\
        CE \& CoT & \textbf{0.671} & \textbf{0.582} & 0.794  & \textbf{0.671} \\ 
        \bottomrule
    \end{tabular}
\end{table}

The modest performance gains from CoT and citation extraction suggest that explicitly structuring the reasoning process through prompting does not meaningfully enhance model performance in statutory retrieval and QA tasks. Although the CoT prompting includes specific guidance for citation processing and negative inference from missing citations (e.g., inferring ``False'' if no citation appeared relevant), this appears to primarily affect response format rather than retrieval and interpretive accuracy. This indicates that statutory retrieval and interpretation require more advanced techniques to handle than CoT and citation extraction.

Despite having minimal effect on performance, citation elicitation with CoT provides us with valuable insight because of the citation analysis done by the inference model on top of the retriever. Here, we prompt for CoT reasoning while also requesting that the inference model return the citation it found most ``relevant'' its answer or\texttt{None} if none are relevant. This citation will be picked out of the top-$k$ retrieved citations (which are sorted by similarity using MIPS as per Section \ref{sec:experiment}). Relevancy, in this case, is not rigidly defined and is determined by the LLM.

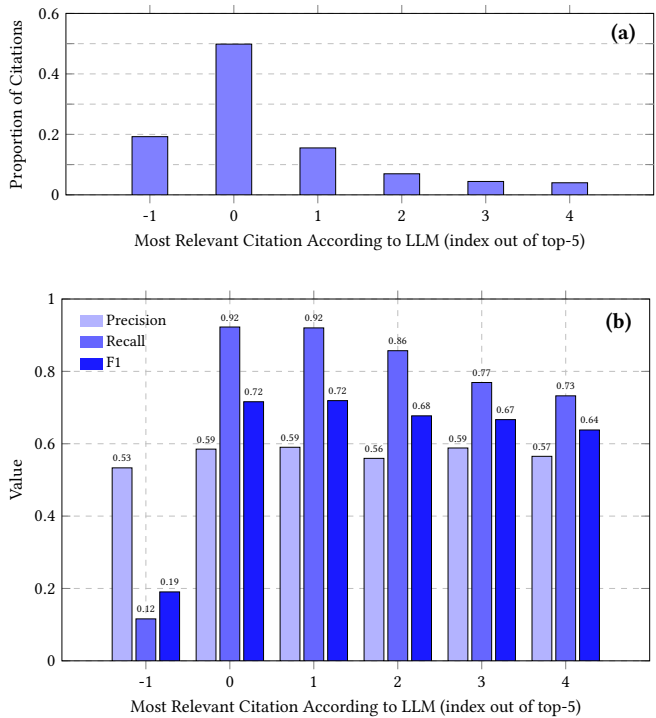
\begin{figure}[!t]
    \centering
    \begin{subfigure}[a]{\linewidth}
        \begin{tikzpicture}[scale=0.75]
\begin{axis}[
    ybar,
    bar width=18pt,
    width=12cm,
    height=4.8cm,
    symbolic x coords={-1, 0, 1, 2, 3, 4},
    xtick=data,
    ymin=0, ymax=0.6,
    xlabel={Most Relevant Citation According to LLM (index out of top-$5$)},
    ylabel={Proportion of Citations},
    enlarge x limits=0.2,
    ymajorgrids=true,
    grid style=dashed,
    ytick={0, 0.2, 0.4, 0.6}, 
    extra y ticks={0.1, 0.3, 0.5}, 
    extra y tick labels={}, 
    extra y tick style={grid style=dashed}, 
]

\addplot[fill=blue!50] coordinates {
    (-1,0.192443)
    (0,0.498516)
    (1,0.155196)
    (2,0.069636)
    (3,0.044265)
    (4,0.039946)
};
\node[anchor=north east, font=\Large\bfseries] at (rel axis cs:0.98,0.98) {(a)};
\end{axis}
\end{tikzpicture}\vspace{-1em}
    \end{subfigure}
    
    \begin{subfigure}[b]{\linewidth}
        \vspace{1em}
       \begin{tikzpicture}[scale=0.75]
\begin{axis}[
    ybar,
    bar width=10pt, 
    width=12cm,
    height=8cm,
    symbolic x coords={-1, 0, 1, 2, 3, 4},
    xtick=data,
    ymin=0, ymax=1,
    xlabel={Most Relevant Citation According to LLM (index out of top-$5$)},
    ylabel={Value},
    enlarge x limits=0.2,
    nodes near coords,
    every node near coord/.append style={font=\tiny},
    grid=major,
    ymajorgrids=true,
    grid style=dashed,
    legend style={
        at={(0.02,0.98)}, 
        anchor=north west, 
        draw=none, 
        fill=none, 
        font=\small 
    },
    legend cell align={left},
    legend image code/.code={
        \draw[fill=#1,draw=none] (0cm,-0.1cm) rectangle (0.4cm,0.1cm);
    }
]

\addplot[fill=blue!30] coordinates {(-1,0.533333) (0,0.585123) (1,0.590244) (2,0.559585) (3,0.588235) (4,0.565217)};
\addplot[fill=blue!60] coordinates {(-1,0.115942) (0,0.922612) (1,0.920152) (2,0.857143) (3,0.769231) (4,0.732394)};
\addplot[fill=blue!90] coordinates {(-1,0.190476) (0,0.716096) (1,0.719168) (2,0.677116) (3,0.666667) (4,0.638037)};

\legend{Precision, Recall, F1}
\node[anchor=north east, font=\Large\bfseries] at (rel axis cs:0.98,0.98) {(b)};
\end{axis}
\end{tikzpicture}


    \end{subfigure}
    \caption{(a) distribution of the citation index produced during inference, representing the position of the returned citation among the top-$k$ results (with 0 as the top result and $-1$ indicating no relevant citations) (b) performance metrics for each index.} \label{fig:citation_index}
\end{figure}

Figure~\ref{fig:citation_index} looks at the index $i$ of the citation deemed most relevant by the LLM out of the top-$5$ fetched during retrieval, where $i=-1$ indicates the model found no results relevant. We observe that the LLM finds the first result ($i=0$) most relevant almost half of the time, and finds no citation relevant almost a fifth of the time. Furthermore, when the LLM finds no citation law, recall is a remarkably low 0.12. A high false negative rate might be expected here as this includes situations where the retriever is unable to fetch the correct statute or regulation. 

\subsubsection{Citation Elicitation Performance.}

We now assess the accuracy of citations elicited. Unfortunately, because DOL has not released citations, this requires us to manually validate answers, which can be a time-consuming process given the complexity of the UI program and its provisions. We hence do this by manually reviewing a simple random sample of 30 outputs from experiments, performed with the GPT, Gemini, and Deepseek generators all using the OpenAI retriever. Each output includes the initial QA from \benchmark, the top-$5$ sections retrieved, the elicited citation---produced through CoT and citation elicitation---and the inferred answer. To evaluate an output, we locate the statute or regulation relevant to the query and review whether the elicited citation or any of the top-$5$ citations match.

We find that the elicited citations have an accuracy rate of 0.73 while OpenAI's retriever achieves a recall@5 of 0.93. The accuracy, conditional on the correct citation appearing in the top-5, is only 0.77. Furthermore, conditional on the elicited citation being correct, the accuracy is 0.79.\footnote{There are situations where the generator will produce the correct True/False answer, but will cite to the wrong section as proof.} 
These results indicate that OpenAI's retriever performs reasonably as an aide for statutory research, but citation accuracy as judged by the (generator) LLM remains limited, even when the appropriate citation was retrieved.

\subsection{State-by-State Results}

One important feature of \benchmark and \corpus's structure is that we are able to compare how the performance of LLMs differs between the state statutory codes. We observe that the states exhibit stark variations in both their baseline knowledge, i.e., without retrieval, and in their knowledge with RAG. 

Analyzing performance on queries made on each state's respective statutory and regulatory code with RAG (averaged across all embeddings and LLMs), F1 scores vary from 0.52  (Virginia) to 0.81 (Arkansas). In a baseline setting, LLM performance varies from 0.316 (Louisiana) to 0.59 (Mississippi). 

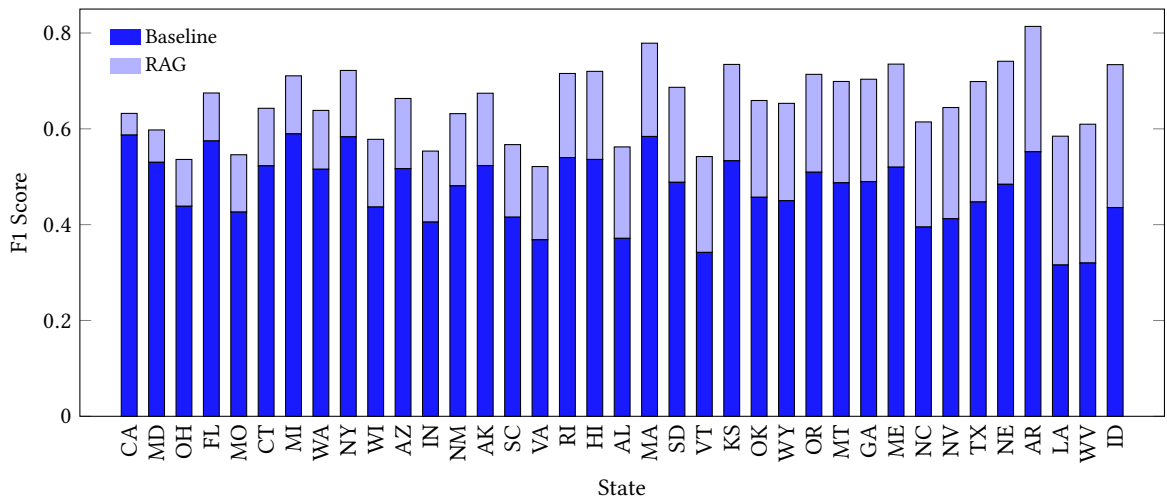
\begin{figure*}[!t]
    \centering
    \begin{tikzpicture}
\begin{axis}[
    ybar stacked,
    bar width=6pt, 
    width=16cm, 
    height=7cm,
    symbolic x coords={CA, MD, OH, FL, MO, CT, MI, WA, NY, WI, AZ, IN, NM, AK, SC, VA, RI, HI, AL, MA, SD, VT, KS, OK, WY, OR, MT, GA, ME, NC, NV, TX, NE, AR, LA, WV, ID},
    xtick=data,
    x tick label style={rotate=90, anchor=east},
    xlabel={State},
    ylabel={F1 Score},
    ymin=0,
    ymax=0.85,
    enlarge x limits=0.05, 
    legend style={
        at={(0.02,0.98)}, 
        anchor=north west, 
        legend columns=1, 
        draw=none, 
        fill=none, 
        font=\small 
    },
    legend cell align={left},
    legend image code/.code={
        \draw[fill=#1,draw=none] (0cm,-0.1cm) rectangle (0.4cm,0.1cm);
    },
    xtick pos=left 
]

\addplot+[ybar, fill=blue!90, draw=black] plot coordinates {
    (CA, 0.586957) (MD, 0.530120) (OH, 0.438356) (FL, 0.574713) (MO, 0.426230) 
    (CT, 0.522727) (MI, 0.589474) (WA, 0.515625) (NY, 0.583333) (WI, 0.436782) 
    (AZ, 0.516854) (IN, 0.405405) (NM, 0.481013) (AK, 0.523364) (SC, 0.415584) 
    (VA, 0.368421) (RI, 0.540000) (HI, 0.536082) (AL, 0.371429) (MA, 0.584071) 
    (SD, 0.488372) (VT, 0.342105) (KS, 0.533333) (OK, 0.457143) (WY, 0.450000) 
    (OR, 0.509434) (MT, 0.487395) (GA, 0.489796) (ME, 0.520000) (NC, 0.395062) 
    (NV, 0.412371) (TX, 0.447368) (NE, 0.484211) (AR, 0.552239) (LA, 0.315789) 
    (WV, 0.320000) (ID, 0.435644)
};
\addlegendentry{Baseline}

\addplot+[ybar, fill=blue!30, draw=black] plot coordinates {
    (CA, 0.045278) (MD, 0.067547) (OH, 0.097742) (FL, 0.100089) (MO, 0.119673) 
    (CT, 0.120159) (MI, 0.121156) (WA, 0.122847) (NY, 0.138533) (WI, 0.141305) 
    (AZ, 0.146518) (IN, 0.148175) (NM, 0.150710) (AK, 0.150935) (SC, 0.151493) 
    (VA, 0.152893) (RI, 0.175543) (HI, 0.183911) (AL, 0.190849) (MA, 0.194696) 
    (SD, 0.198295) (VT, 0.200091) (KS, 0.201107) (OK, 0.201999) (WY, 0.203176) 
    (OR, 0.204281) (MT, 0.211500) (GA, 0.213754) (ME, 0.215088) (NC, 0.219334) 
    (NV, 0.232132) (TX, 0.251355) (NE, 0.256812) (AR, 0.261551) (LA, 0.268867) 
    (WV, 0.289654) (ID, 0.298360)
};
\addlegendentry{RAG}

\end{axis}
\end{tikzpicture}
\vspace{-2.5em}
    \caption{Mean F1 Scores With and Without RAG Across Jurisdictions (Sorted by RAG-Baseline Difference)} \label{fig:rag_vs_no_rag_state}
\end{figure*}

We illustrate our state-by-state RAG and baseline results in Figure \ref{fig:rag_vs_no_rag_state}. There, we sort the states by the difference between their F1 scores with RAG and at baseline. We consider this difference the \textit{knowledge gained from RAG}. We see that the introduction of retrieval itself affects states differently. California receives relatively minimal gains in F1 score ($0.05$) from retrieval, while Idaho benefits considerably ($0.30$ gain in F1 score). Evidently, more work is necessary to understand how LLMs and retrievers handle state codes differently. 

\subsubsection{Correlates with Knowledge Gain From RAG.}
So far, our results suggest that code simplification efforts will benefit in divergent ways across states from AI. This raises the puzzle of why 
there is such diversity in state-level performance. Do differences in legislative processes, administrative organization, or economic demand drive such differential performance of AI systems?  

Prior to inspecting other variables, we examine how population correlates with performance gain with RAG, given population's strong influence on other economic and political factors. We observe a statistically significant negative correlation ($r=-0.40$, $p=0.01$).\footnote{Performing a linear regression, we observe that Texas is an Tukey outlier ($k=1.5$); without Texas, the correlation becomes considerably stronger ($r=-0.58$).}

We study how performance gain with RAG relates to variation in state policy and institutional metrics, obtained from \citet{grossmann2021correlates}. We find no covariates (of 2,970) that have statistically significant predictive power in bivariate tests, adjusting for multiple testing \cite{benjamini1995controlling}. We find similar results in regression analyses. Nonetheless, this combination of institutional and jurisdictional performance may be a fruitful inquiry enabled by \benchmark and \corpus for other comparative work.

\section{Conclusion} \label{sec:conclusion}

Through \benchmark and \corpus, we provide an open framework for evaluating AI system performance in statutory retrieval, comparison, and interpretation. \benchmark and \corpus offer a realistic setting  for assessing  LLMs ability to handle the complexity of statutory language. As calls for regulatory simplification increase pressure for employment of LLMs, robust and open means of evaluating statutory understanding such as \benchmark become increasingly important.

Notwithstanding widely touted efforts to employ AI for ambitious reform efforts, our results reveal that there is considerable room for improvement in the state-of-the-art RAG system's performance in jurisdiction-specific QA. Although using retrieval provides significant gains in F1 and other techniques like CoT provide marginal gains, systems still are unable to ascertain salient dimensions of state UI law. While they may be used as aides for code understanding, current approaches appear to fall short of touted claims that AI can serve as end-to-end solutions for code simplification. 

\subsection{Future Work} First, we anticipate there are many retrieval strategies that may be applied to \corpus to still  increase LLM performance on \benchmark. We expect that many of those strategies that are specific to statutory retrieval and entailment, like those employed in COLIEE tasks 3 and 4 (e.g., \citet{nguyen_captain_2024}'s CAPTAIN) might improve performance on our benchmark. Modifications to prompting and methods such as context generation might also be employed to improve performance without retrieval. Adjusting retrieval granularity is another promising avenue: \citet{lima_unlocking_2024} shows using hierarchical, multi‑layer embeddings mirroring statutory structure---from sections down to individual clauses---could similarly boost \benchmark performance.

Second, we hope the completeness of \corpus as well as its accessible structure enables much richer analysis of state statutory and regulatory law. For reasons articulated above, we focused here on the consequential and legally challenging area of UI law, but researchers can build on \corpus to develop methods across many more domains.\footnote{One limitation is that we were unable to access and structure a small number of states. We will continue to expand \benchmark to include reference citations for each answer when applicable. DOL has historically not included these, but their team appears willing to share this information prospectively. This would allow for testing of both retrieval and entailment separately, and also for testing metrics like recall@$k$.}  

Third, the benchmark can be built out to represent the many years in which the state scan was completed by DOL. Fourth, while we have limited our inquiry to the (already challenging) boolean setting, our benchmark enables evaluation with the wider range of data types (categorical, integer, long and short form text) in \benchmark. 

Last, another fruitful direction would be to test AI-assistance for processes like constructing DOL's annual survey of state UI laws. 

In sum, we hope that \benchmark and \corpus will (1) enable more systematic assessment and tracking of statutory understanding, (2) help advance methods for statutory RAG, and (3) offer a useful resource for research in regulatory domains, including not only the highly consequential domain of unemployment insurance, but beyond. 

\begin{acks}
We are grateful to Neel Guha, Chris Manning, Derek Ouyang, Jen Pahlka, Amy Perez, Faiz Surani, Mirac Suzgun, Lucia Zheng, and participants of the RegLab meeting for comments and feedback. We also thank Faiz Surani for his help and recommendations in applying OCR for optimizing extraction from \textit{CSUIL}.
\end{acks}

\bibliographystyle{ACM-Reference-Format}
\bibliography{references, supplemental}

\end{document}